\documentclass[12pt]{article}
\usepackage{graphicx}

\input{babarsym.tex}

\newcommand\pubdate{\today}

\def\Title#1{\begin{center} {\Large #1 } \end{center}}
\def\Author#1{\begin{center}{ \sc #1} \end{center}}
\def\Address#1{\begin{center}{ \it #1} \end{center}}

\newcommand\pubblock{\rightline{\begin{tabular}{l} %\pubnumber\\
         \pubdate  \end{tabular}}}
\newenvironment{Abstract}{\begin{quotation}  }{\end{quotation}}
\newenvironment{Presented}{\begin{quotation} \begin{center} 
             PRESENTED AT\end{center}\bigskip 
      \begin{center}\begin{large}}{\end{large}\end{center} \end{quotation}}

\def\beq{\begin{equation}}
\def\eeq#1{\label{#1}\end{equation}}
\def\eeqn{\end{equation}}

\def\beqa{\begin{eqnarray}}
\def\eeqa#1{\label{#1}\end{eqnarray}}
\def\eeqan{\end{eqnarray}}

\let\bar=\overbar

\def\Dslash{\not{\hbox{\kern-4pt $D$}}}
\def\dslash{\not{\hbox{\kern-2pt $\del$}}}

\def\BR{\mbox{\rm BR}}

\def\msb{{\bar{\ssstyle M \kern -1pt S}}}

\begin{document}
\begin{titlepage}
\pubblock

\vfill
\Title{CPT, T and CP Violation in B Mixing}
\vfill
\Author{Markus R\"ohrken}
\Address{Charles C. Lauritsen Laboratory of High Energy Physics\\
California Institute of Technology\\1200 East California Boulevard MC 356-48\\CA 91125 Pasadena\\USA}
\vfill
\begin{Abstract}
Recent measurements of observables sensitive to CPT, T and CP violation in neutral B meson mixing by the B-factory experiments Belle and \babar are presented.
\end{Abstract}
\vfill
\begin{Presented}
Flavor Physics \& CP Violation 2014\\
Marseille, France, May 26--30, 2014
\end{Presented}
\vfill
\end{titlepage}
\def\thefootnote{\fnsymbol{footnote}}
\setcounter{footnote}{0}

\section{Introduction}
The B-factory experiments Belle and \babar established CP violation in the neutral and charged B meson system, and experimentally confirmed
the source of CP violation to be one single complex phase in the three-family CKM quark mixing matrix~\cite{Belle2001,BaBar2001,Belle2004c,BaBar2004c}. The Belle and \babar experiments
provide an excellent environment to study heavy-flavor decays. Most importantly the clean environment and the Lorentz boost caused by the asymmetric-energy $e^+e^-$ colliders KEKB and PEP-II,
and the quantum entanglement of neutral B mesons produced by $\Upsilon(4S)$ decays enable measurements of observables sensitive to the fundamental symmetries CP, T and CPT.
At the conference Flavor Physics \& CP Violation 2014 the current most precise measurements sensitive to the breaking of these symmetries have been presented, and compared to previous related results.
In this proceedings article the presented measurements are summarized, and references for further reading are provided.

\section{CPT Measurement by Belle}
The invariance under CPT, the combined operation of charge-conjugation C, parity-transformation P and time-reversal T, is of fundamental importance in physics. All local Lorentz-invariant
quantum field theories are assumed to conserve CPT~\cite{Lueders,Pauli,Greenberg}. A breaking of CPT symmetry would have important consequences such as the possibility for violation of
Lorentz invariance, or differences in lifetimes and masses of particles and corresponding antiparticles. Searches for the violation of CPT symmetry have been carried out before
in the neutral kaon system by the CPLEAR, KLOE and KTeV collaborations~\cite{CPLEAR2001,KLOE2006,KTeV2011}, and in the neutral B meson system by the Belle and \babar experiments~\cite{Belle2003,BaBar2004a,BaBar2006}.

An observable sensitive to CPT violation in neutral B meson mixing is the complex parameter $z$, which is related to the light ($B_{L}$) and heavy ($B_{H}$) mass eigenstates and the complex mixing parameters $p$ and $q$ by:
\begin{eqnarray}
| B_{L} \rangle \propto& p \sqrt{1 - z} | B^{0} \rangle | + q \sqrt{1 + z}  | \bar{B}^{0} \rangle \nonumber \\
| B_{H} \rangle \propto& p \sqrt{1 + z} | B^{0} \rangle | - q \sqrt{1 - z}  | \bar{B}^{0} \rangle \nonumber
\end{eqnarray}

In 2012, Belle performed a measurement of the CPT violating parameter $z$ based on a data sample containing $535 \times 10^6$ $B\bar{B}$ pairs collected on the $\Upsilon(4S)$~\cite{Belle2012}.
The time-dependent analysis utilizes the coherent mixing of two entangled neutral B mesons in an $\Upsilon(4S)$ event~\cite{BaBar2004a,BaBar2004b}. The measurement reconstructs one B meson either in CP eigenstates,
or in flavor-specific hadronic or semileptonic decay modes. The reconstructed CP modes $B^0 \to J/\psi K^{0}_{S}$ and $B^0 \to J/\psi K^{0}_{L}$ provide sensitivity mainly on $Re(z)$, while others are sensitive to $Im(z)$.
Besides the measurement of $Re(z)$ and $Im(z)$, the analysis extracts $\Delta \Gamma_{d} / \Gamma_{d}$ from an unbinned maximum likelihood fit of the reconstructed data containing about $560k$ events
to the full decay rate of the entangled B meson pairs in an $\Upsilon(4S)$ event. The obtained results are:

\begin{eqnarray}
Re ( z ) =& \left[ +1.9 \pm 3.7 \, (\rm{stat}) \pm 3.3 \, (\rm{syst}) \right] \times 10^{-2} \nonumber \\
Im ( z ) =& \left[ -5.7 \pm 3.3 \, (\rm{stat}) \pm 3.3 \, (\rm{syst}) \right] \times 10^{-3} \nonumber \\
\frac{\Delta \Gamma_{d}}{\Gamma_{d}} =& \left[ -1.7 \pm 1.8 \, (\rm{stat}) \pm 1.1 \, (\rm{syst}) \right] \times 10^{-2} \nonumber
\end{eqnarray}

The results are consistent with the assumption of no CPT violation, and provide the current most precise estimation of observables sensitive to CPT violation in the neutral B meson system.

\section{T Violation Measurement by \babar}
To compensate for CP violating effects the invariance under CPT of Standard Model physics processes predicts T violating phenomena.
To observe T violation by rate comparisons, initial and final states of physical processes need to be exchanged.
Performing this exchange by the reversal of reactions of unstable particles is difficult to realize in experiments. In 2012, \babar performed a measurement utilizing the coherent mixing of
two neutral B mesons in an $\Upsilon(4S)$ event 
to probe for time-reversal invariance by measuring the rate differences of processes with exchanged initial and final states~\cite{BaBar2012}. 

The measurement follows an idea by Banuls and Bernabeu~\cite{Banuls1999, Banuls2000} and applies a procedure called CP tagging in analogy to the flavor-tagging in standard time-dependent CP violation measurements.
This enables the inference of the CP state of one B meson at the instant of decay of the second B meson.
The approach allows the construction of proper states and the measurement of the rates of reactions related by T transformation.
For example, the transition of $\bar{B}^{0} \to B_{-}$ where $B_{-}$ denotes a CP-odd eigenstate is identified by the time-ordered final states $(l^{+}, J/\psi K^{0}_{S})$.
The related T reversed process $B_{-} \to \bar{B}^{0}$ is obtained by the final states $(J/\psi K^{0}_{L}, l^{-})$.
The measurement provides three further independent rate comparisons between $B_{+} \to B^{0}$ $(J/\psi K^{0}_{S}, l^{+})$, $\bar{B}_{0} \to B_{+}$ $(l^{+}, J/\psi K^{0}_{L})$, and 
$B_{-} \to B^{0}$ $(J/\psi K^{0}_{L}, l^{+})$ and their corresponding T-conjugated transitions. Any difference in the rates of these pairs manifests the breaking of T symmetry.
Furthermore the measurement needs no assumptions about CP violation or CPT conservation as input, but measures asymmetry parameters sensitive to CP and CPT in addition to that sensitive to T violation.

For the main T violating parameters, \babar measures using a data sample containing $468 \times 10^6$ $B\bar{B}$ pairs collected on the $\Upsilon(4S)$:

\begin{eqnarray}
\Delta S^{+}_{T} =& -1.37 \pm 0.14 \, (\rm{stat}) \pm 0.06 \, (\rm{syst}) \nonumber \\
\Delta S^{-}_{T} =& 1.17 \pm 0.18 \, (\rm{stat}) \pm 0.11 \, (\rm{syst}) \nonumber
\end{eqnarray}

The result directly observes T violation with a significance of $14\sigma$, and is in agreement with the assumption of CPT conservation and CP violation.
 
\begin{figure}[htb]
\centering
\includegraphics[width=0.7\textwidth]{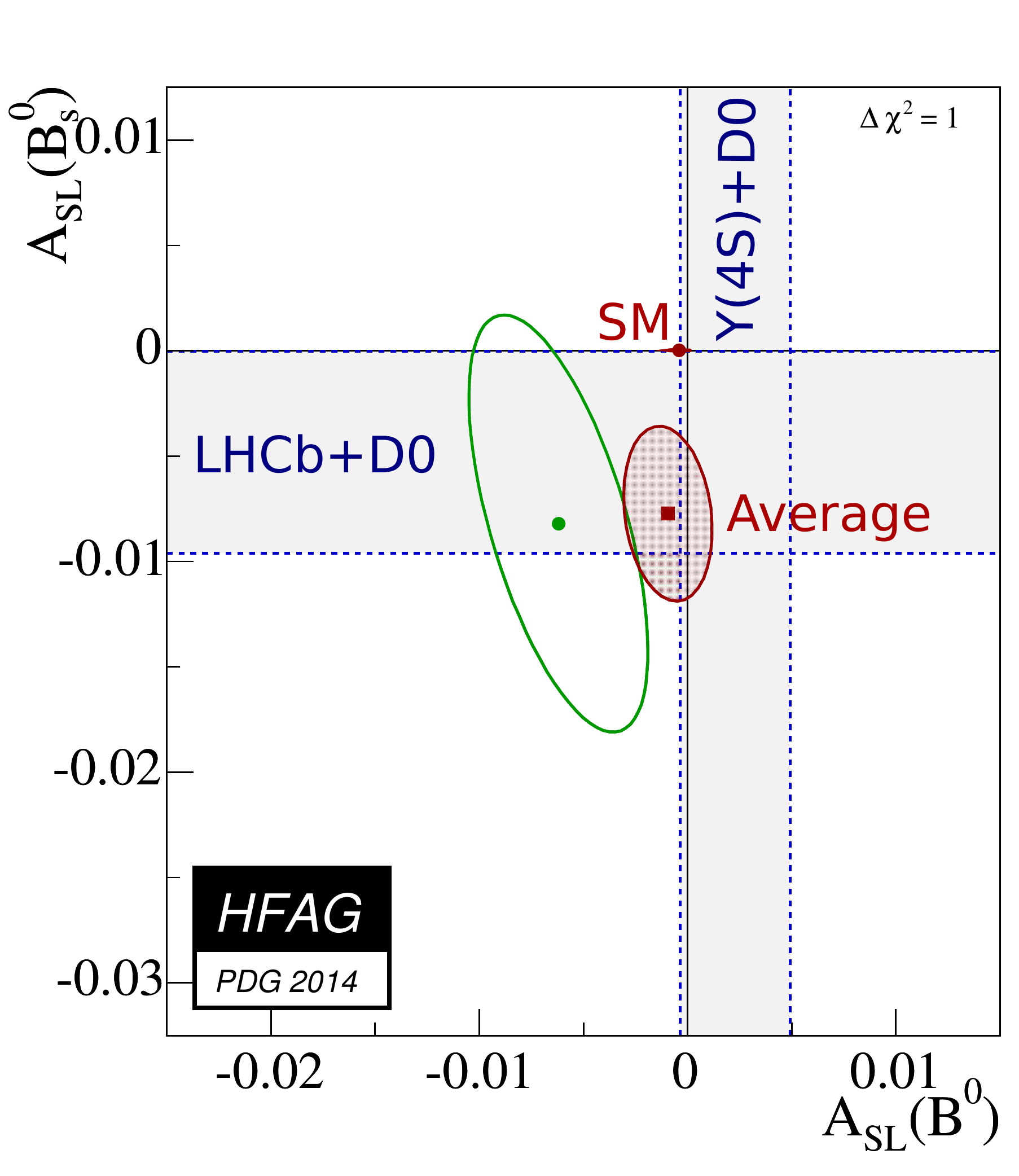}
\caption{Comparison and average of $A_{SL}^{s}$ and $A_{SL}^{d}$ measurements provided by the Heavy Flavor Averaging Group (HFAG)~\cite{HFAG}.}
\label{fig:HFAG_average}
\end{figure} 
 
\section{CP Violation in Mixing Measurement by \babar}

CP violation in neutral B meson mixing provides a sensitive probe for models beyond the Standard Model. New particles might emerge in the processes mediated by box diagrams and modify the mixing asymmetry defined by~\cite{Lenz2007}:
\[
 A_{SL}^{q} = \frac{ P \left( \bar{B}^{0}_{q} \to B^{0}_{q} \left(t\right) \right) - P \left( B^{0}_{q} \to \bar{B}^{0}_{q} \left(t\right) \right) }{ P \left( \bar{B}^{0}_{q} \to B^{0}_{q} \left(t\right) \right) + P \left( B^{0}_{q} \to \bar{B}^{0}_{q} \left(t\right) \right) }
= \frac{1 - |q/p|^{4}}{1 + |q/p|^{4}} \approx \frac{|\Gamma_{12}^{q}|}{|M_{12}^{q}|} \sin \phi_{q}
\]
The Standard Model predictions for the neutral $B_d$ meson system are $A_{SL}^{d} = (1.8 \pm 0.3) \times 10^{-5}$ and $\phi_{d} = (0.24 \pm 0.06)^{\circ}$~\cite{Nierste2012}. The current achievable experimental sensitivity
on $A_{SL}^{q}$ is $\mathcal{O}(10^{-3})$. Any significant deviation from 0 might point to possible beyond the Standard Model processes. An important motivation for measurements sensitive to CP violation in B mixing comes from the tension driven
by the results of the D0 experiment on the inclusive like-sign dimuon charge asymmetry~\cite{Dzero2014}.

In 2013, \babar performed a search for CP violation in $B^{0}$-$\bar{B}^{0}$ mixing using a data sample containing $468 \times 10^6$ $B\bar{B}$ pairs collected on the $\Upsilon(4S)$~\cite{BaBar2013}.
The measurement reconstructs one B meson as $B^{0} \to D^{*-} X l^{+} \nu$ by applying a partial reconstruction technique for $D^{0}$ mesons from $D^{*-} \to D^{0} \pi^{-}_{\mathrm{soft}}$ decays.
The flavor of the reconstructed B meson is inferred from the lepton charge, and the flavor of the second B meson is inferred from its decay products by employing a kaon tag.
A difficulty of the measurement is the correction for effects induced by physical and detector charge asymmetries. The strategy of the measurement is to avoid relying on control samples or Monte Carlo simulations, and to
estimate these effects directly from data by exploiting all available information from different reconstructed sub-samples accounting for electron or muon leptons, and mixed or unmixed neutral B events.
The signal fraction is estimated from a fit to distributions of the missing neutrino mass, and a yield of $(5.945 \pm 0.007) \times 10^6$ signal events obtained. The $A_{SL}^{d}$ asymmetry is obtained
by a binned four-dimensional fit to the $\Delta t$, $\sigma(\Delta t)$, $\cos(\theta_{lk})$ and $p_k$ distributions. The result of the measurement is:

\begin{eqnarray}
| \frac{q}{p} - 1 | =& ( -0.29 \pm 0.84 \, (\rm{stat}) \, ^{+1.88}_{-1.61} \, (\rm{syst}) ) \times 10^{-3} \nonumber \\
A_{SL}^{d} =& ( 0.06 \pm 0.17 \, (\rm{stat}) \, ^{+0.38}_{-0.32} \, (\rm{syst}) ) \% \nonumber \nonumber
\end{eqnarray}

The measurement provides the most precise single result on $A_{SL}^{d}$, and is in agreement with the Standard Model predictions. A comparison of results from $e^+e^-$ and hadron colliders provided by the Heavy Flavor Averaging Group is shown in Figure~\ref{fig:HFAG_average}.

%%%%%%%%%%%%%%%%%%%%%%%%%%%%%%%%%%%%%%%%%%%%%%%%%%%%%%%%%%%%%%%%%%%%%%%%%
%%
%%   use this format to include an .eps figure into your paper
%%
% \begin{figure}[htb]
% \centering
% \includegraphics[height=1.5in]{magnet}
% \caption{Plan of the magnet used in the mesmeric studies.}
% \label{fig:magnet}
% \end{figure}
%%%%%%%%%%%%%%%%%%%%%%%%%%%%%%%%%%%%%%%%%%%%%%%%%%%%%%%%%%%%%%%%%%%%%%%%%%%


\begin{thebibliography}{99}

%%
%%  bibliographic items can be constructed using the LaTeX format in SPIRES:
%%    see    http://www.slac.stanford.edu/spires/hep/latex.html
%%  SPIRES will also supply the CITATION line information; please include it.
%%

\bibitem{Belle2001}
K. Abe et al. (Belle Collaboration), Phys. Rev. Lett. {\bf 87}, 091802 (2001).

\bibitem{BaBar2001}
B. Aubert et al. (\babar Collaboration), Phys. Rev. Lett. {\bf 87}, 091801 (2001).

\bibitem{Belle2004c}
Y. Chao et al. (Belle Collaboration), Phys. Rev. Lett. {\bf 93}, 191802 (2004).

\bibitem{BaBar2004c}
B. Aubert et al. (\babar Collaboration), Phys. Rev. Lett. {\bf 93}, 131801 (2004).

\bibitem{Lueders}
G. L\"uders, Math. Fysik. Medd. Kgl. Danske Akad. Ved. {\bf 28}, 5 (1954).

\bibitem{Pauli}
W. Pauli, ``Niels Bohr and the Development of Physics'', McGraw-Hill, 1st edition, pp. 30--51 (1955).

\bibitem{Greenberg}
O. W. Greenberg, Found. Phys. {\bf 36}, 1535 (2006).

% CPT measurements in kaon system
\bibitem{CPLEAR2001}
A. Angelopoulos et al. (CPLEAR Collaboration), Eur. Phys. J. C {\bf 22}, 55 (2001).

\bibitem{KLOE2006}
F. Ambrosino et al. (KLOE Collaboration), J. High Energy Phys. {\bf 12}, 011 (2006).

\bibitem{KTeV2011}
E. Abouzaid et al. (KTeV Collaboration), Phys. Rev. D {\bf 83}, 092001 (2011).

% CPT in B meson system
\bibitem{Belle2003}
N. C. Hastings et al. (Belle Collaboration), Phys. Rev. D {\bf 67}, 052004 (2003).

\bibitem{BaBar2004a}
B. Aubert et al. (\babar Collaboration), Phys. Rev. D {\bf 70}, 012007 (2004).

\bibitem{BaBar2006}
B. Aubert et al. (\babar Collaboration), Phys. Rev. Lett. {\bf 96}, 251802 (2006).

\bibitem{Belle2012}
T. Higuchi et al. (Belle Collaboration), Phys. Rev. D {\bf 85}, 071105 (2012).

\bibitem{BaBar2004b}
B. Aubert et al. (\babar Collaboration), Phys. Rev. Lett. {\bf 96}, 181801 (2004).

% T violation
\bibitem{BaBar2012}
J. P. Lees et al. (\babar Collaboration), Phys. Rev. Lett. {\bf 109}, 211801 (2012).

\bibitem{Banuls1999}
M. C. Banuls, J. Bernabeu, Phys. Lett. B {\bf 464}, 117 (1999).

\bibitem{Banuls2000}
M. C. Banuls, J. Bernabeu, Nucl. Phys. B {\bf 590}, 19 (2000).

\bibitem{Lenz2007}
A. Lenz, U. Nierste, J. High Energy Phys. {\bf 06}, 72 (2007).

\bibitem{Nierste2012}
U. Nierste, arXiv:1212.5805.

\bibitem{Dzero2014}
V. M. Abazov et al. (D0 Collaboration), Phys. Rev. D {\bf 89}, 012002 (2014).

\bibitem{BaBar2013}
J. P. Lees et al. (\babar Collaboration), Phys. Rev. Lett. {\bf 111}, 101802 (2013).

\bibitem{HFAG}
D. Asner et al. (Heavy Flavor Averaging Group), arXiv:1010.1589.

\end{thebibliography}
\end{document}